\documentclass[preprint,5p]{elsarticle}

\usepackage{graphicx}
\usepackage{dcolumn}
\usepackage{bm}
\usepackage{color}
\usepackage{amsmath,amssymb,amsthm,color}

\begin{document}


\title{Conformal Behavior in QCD}

\author[KI]{K.-I. Ishikawa}
\address[KI]{Graduate School of Science, Hiroshima University,Higashi-Hiroshima, Hiroshima 739-8526, Japan}

\author[YI]{Y. Iwasaki}
\address[YI]{Center for Computational Sciences, University of Tsukuba,Tsukuba, Ibaraki 305-8577, Japan}

\author[YN]{Yu Nakayama}
\address[YN]{California Institute of Technology,  Pasadena, CA 91125, USA}

\author[TY]{T. Yoshie}
\address[TY]{Center for Computational Sciences, University of Tsukuba,Tsukuba, Ibaraki 305-8577, Japan}

\date{\today}

\begin{abstract}
We give a new perspective on the properties of quarks and gluons at finite temperature $T$
in $N_f=2\sim 6 $ QCD.
We point out the existence of an IR fixed point for the gauge coupling constant at $T>T_c$ ($T_c$ is the chiral phase transition temperature).
Based on this observation
we predict theoretically that 
the propagators of a meson $G_H(t)$ at $T/T_c > 1$ decay with a power-law corrected Yukawa-type decaying form, $G_H(t)=c\, \exp{(-\tilde{m}_H t)}/t^{\alpha}$,
{instead of the exponentially decaying form $c_H\exp{(-m_H t)}$,}
 in the ''conformal region'' defined by $m_H < c\, \Lambda_{\mathrm{IR}}$,
where $\Lambda_{\mathrm{IR}}$ is the IR cutoff,  $m_H, \tilde{m}_H$ are characteristic scales of the spectrum in the cannel $H$ 
and $c$ is a constant of order $1$.
The decaying form is the characteristics of conformal theories with an IR cutoff.
This prediction is also applicable to any QCD with compact space in the deconfining region.
We verify numerically  the conjecture on a lattice with size $16^3\times 64$.
We discuss in detail how the resulting hyper-scaling relation of physical  observables may modify the existing argument about the order of the chiral phase transition in the $N_f=2$ case.
\end{abstract}


\maketitle

The properties of quarks and gluons at high temperature are key ingredients for understanding the
evolution of the Universe and the heavy ion collision experiment.
Lattice QCD is the most reliable formulation of QCD for the investigation of  non-perturbative properties of
quarks and gluons, and indeed a lot of works from the early stage of lattice gauge theories were devoted to clarify them~\cite{pioneer}.
Although many interesting and useful results were obtained,
still there remain unsolved issues. 
Among other things, it is a long standing issue to determine the order of the chiral transition for $N_f=2$ QCD, which has recently attracted a lot of attention~\cite{review}.

In this article we give a new perspective on the properties of
quark-gluon state above the chiral phase transition temperature $T_c,$
and thereby give a new insight into the order of the chiral phase
transition. Our main idea is to use the concept of the ''conformal field
theory with an IR cutoff''.
After presenting the renormalization group
argument and conjecturing that the propagator decays with a power-law
corrected Yukawa-type decaying form, we discuss numerical results of
lattice calculations and their physical implications on the chiral
transition in $N_f=2$ QCD.
Some preliminary results have been presented in \cite{iwa2012}.

We discuss lattice QCD at high temperature
for small $N_f$ ($2 \le N_f \le 6$)  massless fermions in fundamental representation where the chiral phase transition occurs at some critical temperature $T_c$.
Our general argument that follows can be applied to any number of flavors ($2 \le N_f \le 6$)
with any formulation of 
gauge theories on the lattice. 

For numerical simulations in this article we take $N_f=2$ and employ the Wilson quark action and the standard one-plaquette gauge action
on the Euclidean lattice of the size $N_x=N_y=N_z=N$ and $N_t,$ 
with the lattice spacing $a.$
We impose an anti-periodic boundary condition in the time direction for fermion fields
and periodic boundary conditions otherwise.
In order to obtain physical quantities at temperature $T$,
we have to take the thermodynamic limit $N \rightarrow \infty$  and the continuum limit $a \rightarrow 0$,
keeping $N_t\, a =1/T$ fixed. 
When the space is compact, $N \, a =L$ is also fixed finite.

The theory is defined by two parameters; the bare coupling constant $g_0$ and the bare degenerate quark mass $m_0$ at ultraviolet (UV) cutoff.
We also use, instead of $g_0$ and $m_0$, 
$\beta={6}/{g_0^2}$
and 
$K= 1/2(m_0a+4)$. 
We define the quark mass $m_q$
through Ward-Takahashi identities
\cite{Bo, ItohNP}
with renormalization constants being suppressed.
The quark mass $m_q$ does only depend on $\beta$ and $K$ and does not depend on whether it is in the confining region or the deconfining region up to order $a$ corrections\cite{iwa_chiral}.

In addition to them 
we investigate in detail  the $t$ dependence of the propagator of the local meson operator in the $H$ channel:
\begin{equation}
G_H(t) = \sum_{x} \langle \bar{\psi}\gamma_H \psi(x,t) \bar{\psi} \gamma_H \psi(0) \rangle \ ,
\label{propagator}
\end{equation}
where the summation is over the spatial lattice points.
In this paper, we mostly focus on the pseudo-scalar (PS) channel $H=PS$. The quark masses and meson masses are expressed in units of the inverse of the lattice spacing $a^{-1}$ in the text and the figures.

We first consider the case where the renormalized quark mass is zero.
Then the renormalized coupling constant is the only relevant variable in the theory. 
A running coupling constant $g(\mu; T)$  at temperature $T$ can be defined as in the case of $T=0$.
Several ways to define the running coupling constant $g(\mu; T)$ are proposed in the 
 literatures (see e.g. ref. \cite{karsch}). 

In the UV regime, since the theory is asymptotically free, 
the running coupling constant at finite $T$ can be expressed as a power series of the running coupling constant at $T=0$ as long as  $g$ is small~\cite{step_scaling}. 
The leading term is universal in the limit $g \rightarrow 0$.

However, in the IR region, the $\mu$ dependence of $g(\mu; T)$ is quite different from that of $g(\mu;T=0)$, since the IR cutoff $\Lambda_{\mathrm{IR}}$ in the time direction is $T$, 
while the IR cutoff is zero at zero temperature. When $\mu$ approaches to the IR cutoff, the difference will be physically significant.

There are some arbitrariness in defining the running coupling constant  $g(\mu; T)$  in the IR region, as in the case $g(\mu;T=0)$. The $\mu$ dependence of $g(\mu; T)$ corresponds to a response to the RG transformation. Thus, as in the case at zero temperature, the arbitrariness comes from the arbitrariness of RG transformation. The scale transformation towards the IR region should integrate properly the degree of freedom in the UV region, in such a way that the  resulting theory be an effective theory in the IR region.
Otherwise, the beta function could possess a fake fixed point. 

Let us quickly recall the running of the coupling constant in QCD at $T=0$. In the UV limit $\mu \rightarrow \infty$, the coupling constant behaves as $g(\mu) \rightarrow 0$ reflecting the asymptotic freedom. As the running scale $\mu$ decreases, $g(\mu)$ increases and at $\mu \simeq \Lambda_{QCD}$, $g(\mu)$ rapidly increases toward $g(\mu) \to \infty$.

Next consider the running of the coupling constant $g(\mu;T)$ at  a fixed finite temperature $T$. The RG transformation does stop evolving at the IR cutoff $T$.
As far as $T < T_c$,  the $g(\mu)$ has evolved already large enough at the scale $T$ and therefore the IR behavior is  in the ``confining region''.  The system is qualitatively similar to that at $T=0$.
On the other hand when $T>T_c$, the $g(\mu;T)$ is in the middle of the process of evolving toward infinity but stop evolving at the scale $T$, as shown in Fig.1. 
Thus the system at $T>T_c$ is quite different from that at $T<T_c$. This is the chiral phase transition, which occurs at $T \simeq \Lambda_{QCD}$.

We therefore claim that there exists an IR fixed point at $\mu \simeq T$ when $T/T_c >1$.
When $T \sim T_c$ with $T > T_c$, the IR fixed point is located at large $g$.  As temperature increases the fixed point moves toward smaller $g$. 

\begin{figure}[thb]
\includegraphics[width=8.0cm]{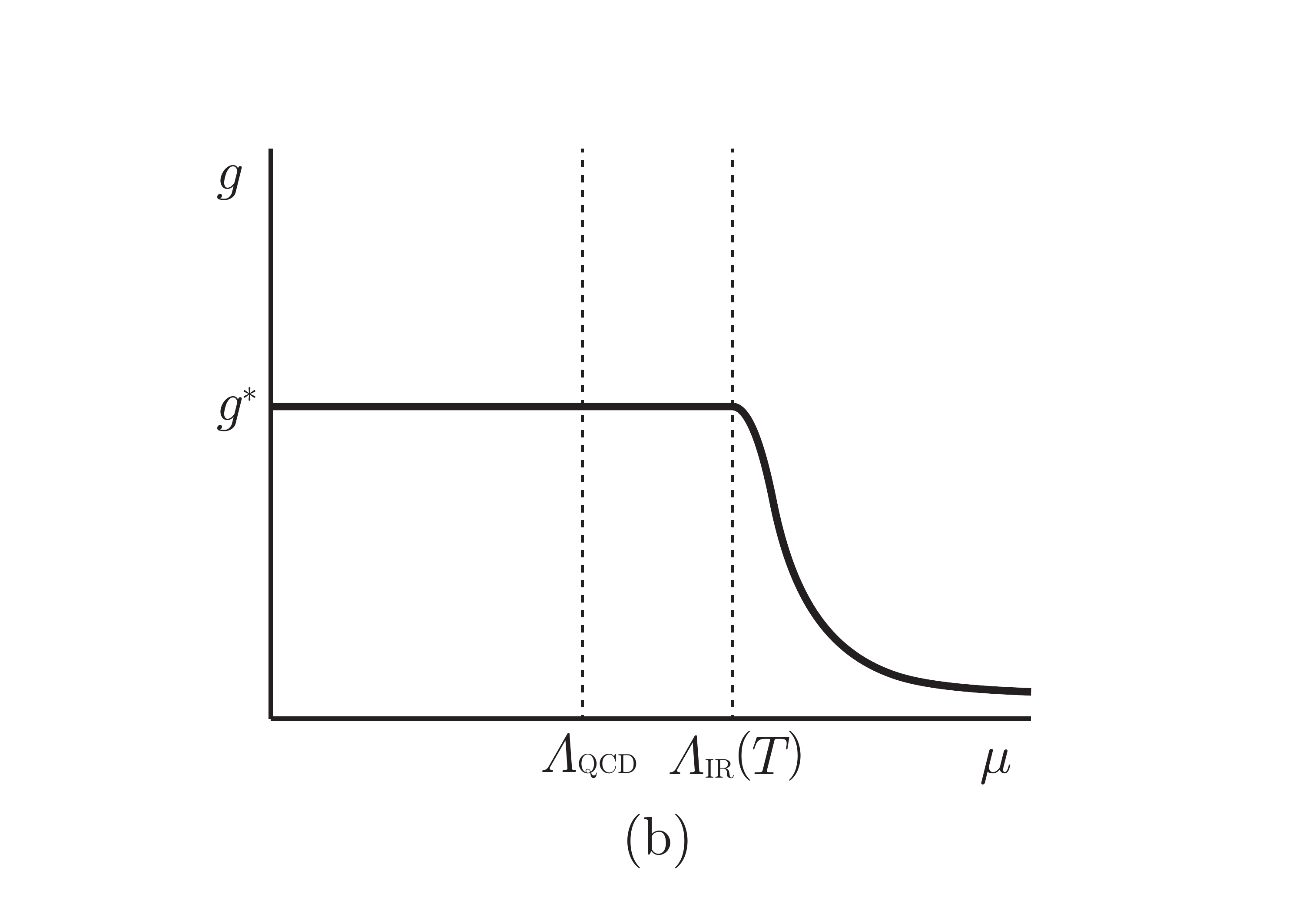}
\caption{
The running of the coupling constant $g(\mu)$ when the $T \simeq \Lambda_{IR}$ is larger than $\Lambda_{QCD}$}
\label{beta_gT}
\end{figure}

Now let us consider the case where the renormalized quark mass is non-zero.
When the typical mass scale (e.g. that of a meson) $m_H$ at the IR cutoff is smaller than $\Lambda_{IR}$, it is in the ``conformal region''.
On the other hand, when $m_H$ is larger than $\Lambda_{\mathrm{IR}}$, the RG flow passes  away from the IR fixed point to a point in the deconfining region with relevant variables integrated out, thus being in the  ``deconfining region''. 
The existence of the conformal region was first discussed in the paper \cite{coll} in the study of many flavor conformal gauge theories at zero temperature. At zero temperature, without the IR cutoff, non-zero quark mass would imply that the theory is in the confining phase (even if $N_f$ is in the conformal window). The reason why the conformal region exists even with the finite quark mass is that the finite IR cutoff makes the running coupling constant stop evolving in parallel with the finite temperature situation discussed above.
One of our goals in this article is to generalize the story to the finite temperature.

Note that this scenario implies that
when physical quantities at IR (e.g. hadron masses) are mapped into a diagram in terms of physical parameters at UV (e.g. the bare coupling constant and the bare quark mass), there will be gaps in the physical quantities along the boundary between the two phases.
We predict the phase transition will be a first order transition there. 

This is a simple, but important observation which is the key in this article. 
We point out two remarks concerning  the observation. 

The one is the relation with the vanishing of the beta function and the energy-momentum trace anomaly.
We recall the relation between the trace anomaly of energy momentum tensor and the beta function with massless quarks: 
\begin{align}
\langle T^{\mu}_{\ \mu} \rangle|_T = \beta(g^{-2}(\mu))  \langle \mathrm{Tr}(F_{\mu\nu} (\mu))^2 \rangle|_T \ , 
\label{equiv}
\end{align}
where $\beta(g^{-2}(\mu))$ is the zero temperature beta function evaluated at $g = g(\mu)$, and 
$\mathrm{Tr}(F_{\mu\nu} (\mu))^2 \rangle|_T$ is the field strength squared at temperature $T$ renormalized at scale $\mu$. 
%
%

\begin{figure*}[thb]
   \includegraphics[width=7.5cm]{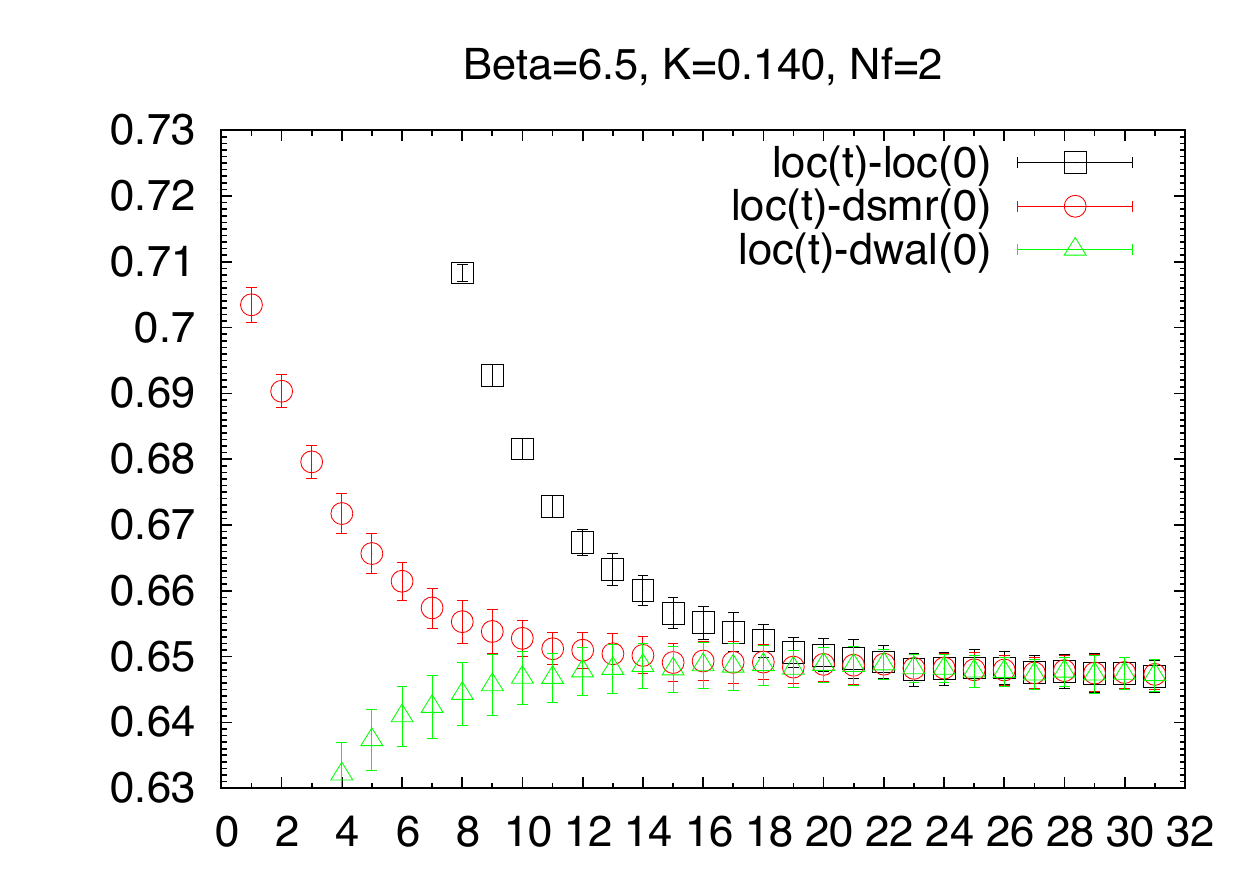}
          \hspace{1cm}
   \includegraphics[width=7.5cm]{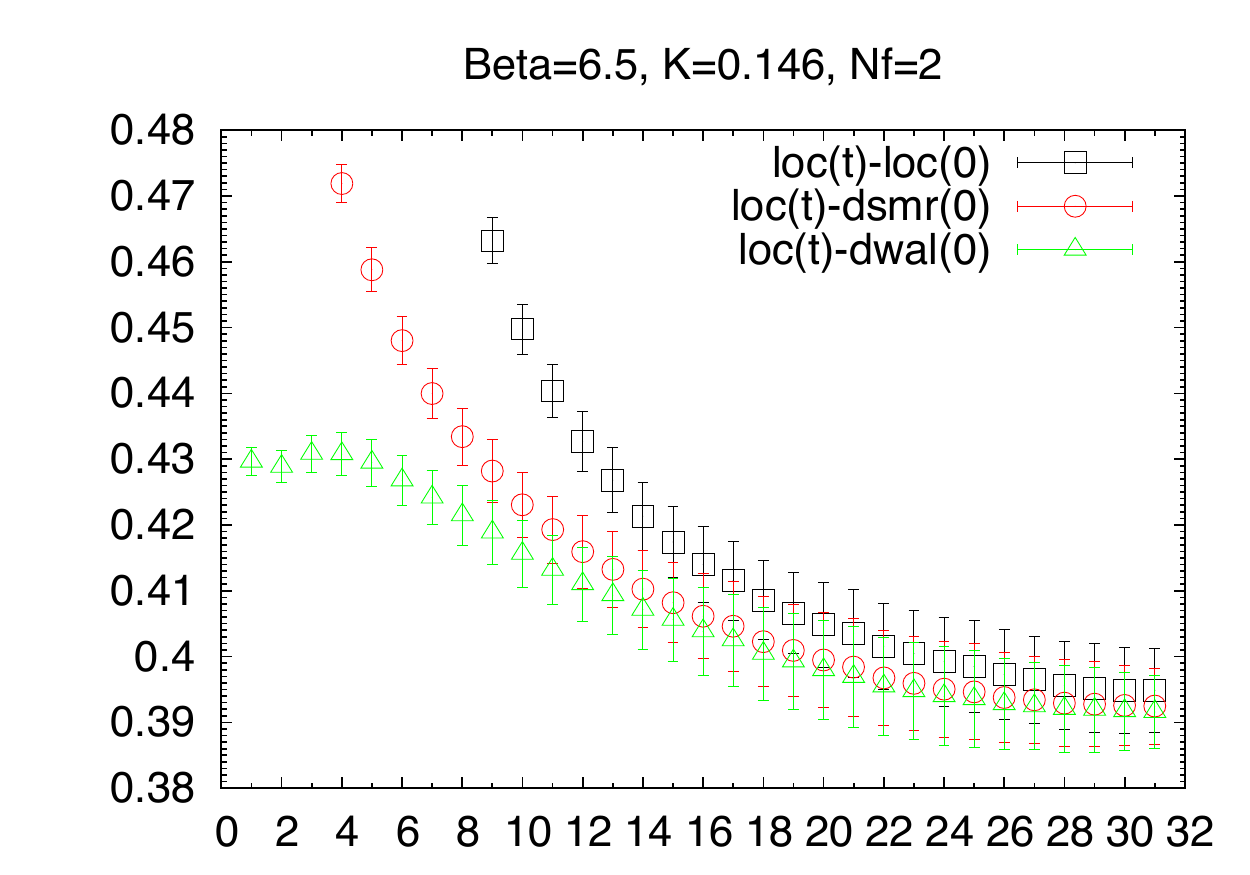}
\caption{(color online).
The effective mass  at $\beta=6.5$ (left: for $K=0.140$ and right: for $K=0.146$).
}
 \label{exp-yukawa}
\end{figure*}

In Lorentz invariant zero-temperature field theories, the vanishing
beta function means that the trace anomaly vanishes
and the theory is conformal invariant in the chiral limit $m_q = 0$ (see e.g. \cite{Nakayama:2013is} for a review). 
In our situation, however, we claim that the beta
function at finite temperatures vanishes,
 which does not  imply vanishing of the trace  of the energy-momentum tensor.
 Thus the vanishing beta function at $T>T_c$ does not contradict with the non-vanishing of
the difference of energy density and three times the pressure.
The other is that at finite temperature $T$, the Euclidean temporal circle plays a role of an {\it{intrinsic}} IR cutoff.
Therefore theories at $T/T_c > 1$  are examples of ''conformal theories with an IR cutoff'', which was introduced
 in \cite{coll}.
We note that QCD at finite temperature even with the massless quark never exhibit power behavior of the meson propagators, because of the {\it{intrinsic}} IR cutoff, despite of the existence of the IR fixed point. 

Verification of the conformal theories with an IR cutoff in QCD at $T/T_c >1$ would make a huge impact.
First of all, it will imply the existence of the conformal region in addition of the confining region and the deconfining region in the phase space as shown below.
It will further provide a fundamental basis  to investigate non-perturbative properties of quarks and gluons at high temperature, such as the slow approach of  the free energy to the Stefan-Boltzmann ideal gas limit.
It also implies, as we discuss below,
non-analytic behavior of the  $m_{PS}$ in terms of the $m_q$, which may be a solution of recent 
issue of the order of the chiral phase transition for $N_f=2$. 

We stress that QCD in compact space 
is also a conformal theory with an IR cutoff for $\beta \ge \beta^c$. Here the $\beta^c$ is the chiral transition $\beta$.
In the case of the compact space, the temperature may be defined by $1/N_t \, a$ as usual.
It may be sated simply that QCD with small mass quarks in the deconfining region is a conformal theory with an IR cutoff.
We also note that
all numerical simulations on a finite lattice for $\beta \ge \beta^c$ belong to conformal theories with an IR cutoff.

\begin{figure*}[thb]
  \hspace{0.5cm}
   \includegraphics[width=7.5cm]{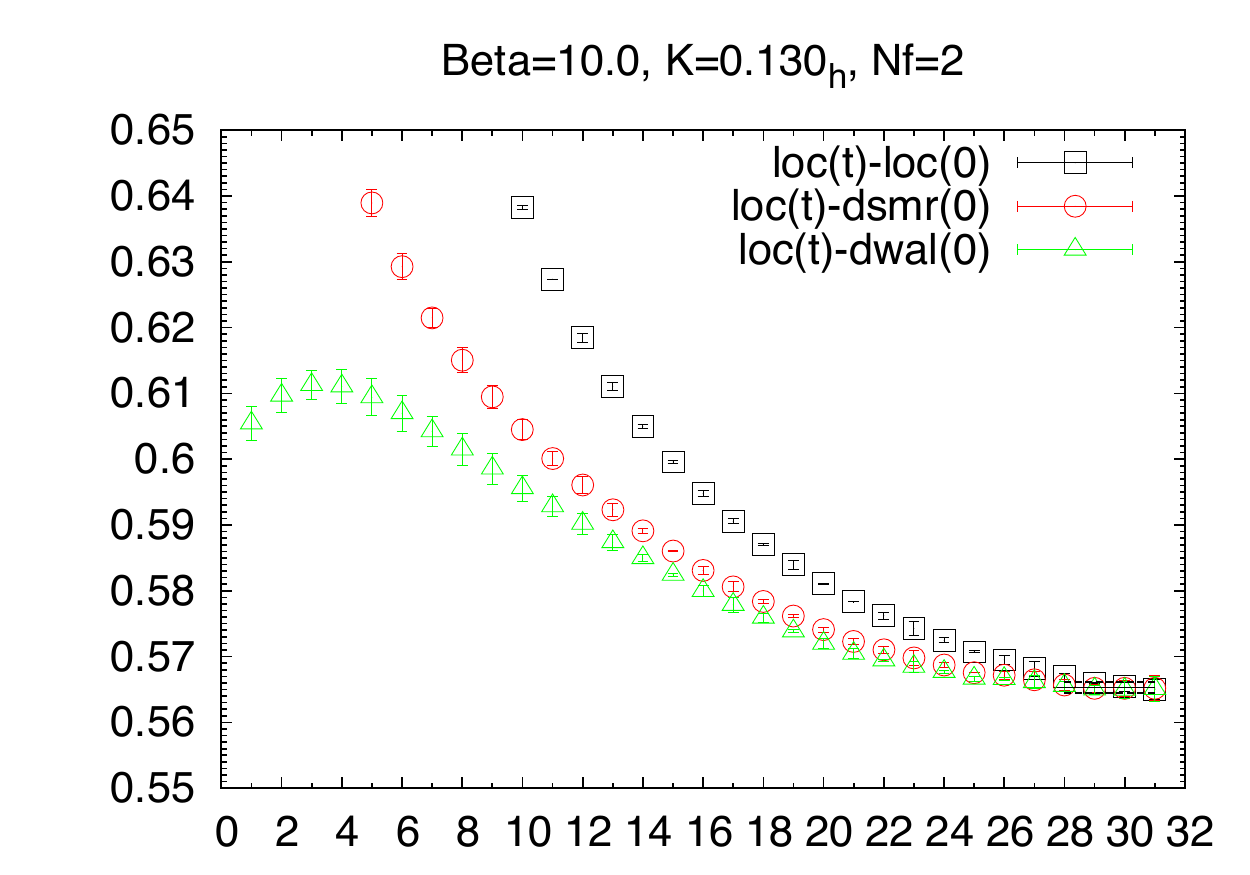}
    \hspace{1cm}
      \includegraphics[width=7.5cm]{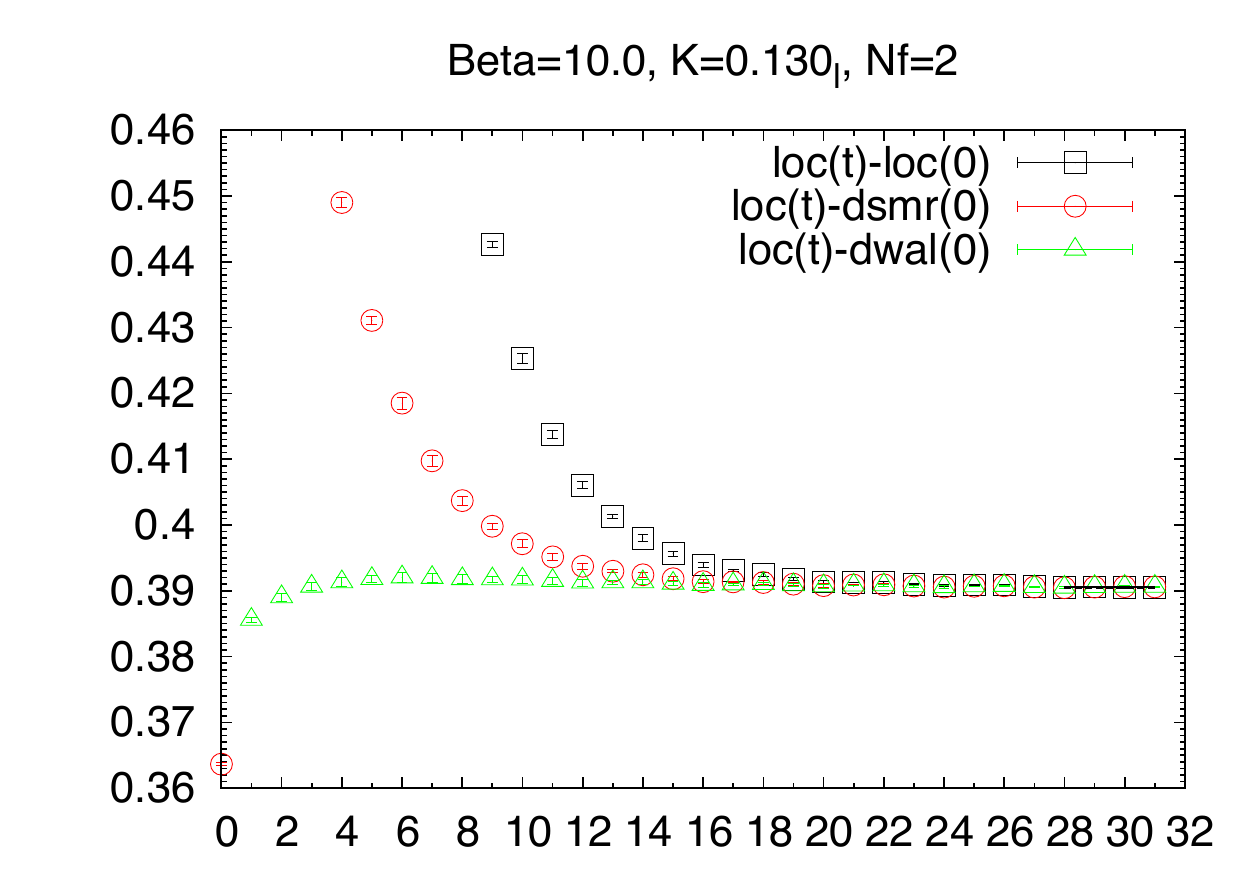}
                       \caption{(color online). The effective mass of two states at the same parameter; $K=0.130$ and $\beta=10.0$:
                        (left: continuation from larger$K$ and right: continuation from smaller $K$. }
                        \label{effm_k130}
\end{figure*}

\begin{figure*}[thb]
   \includegraphics[width=8.0cm]{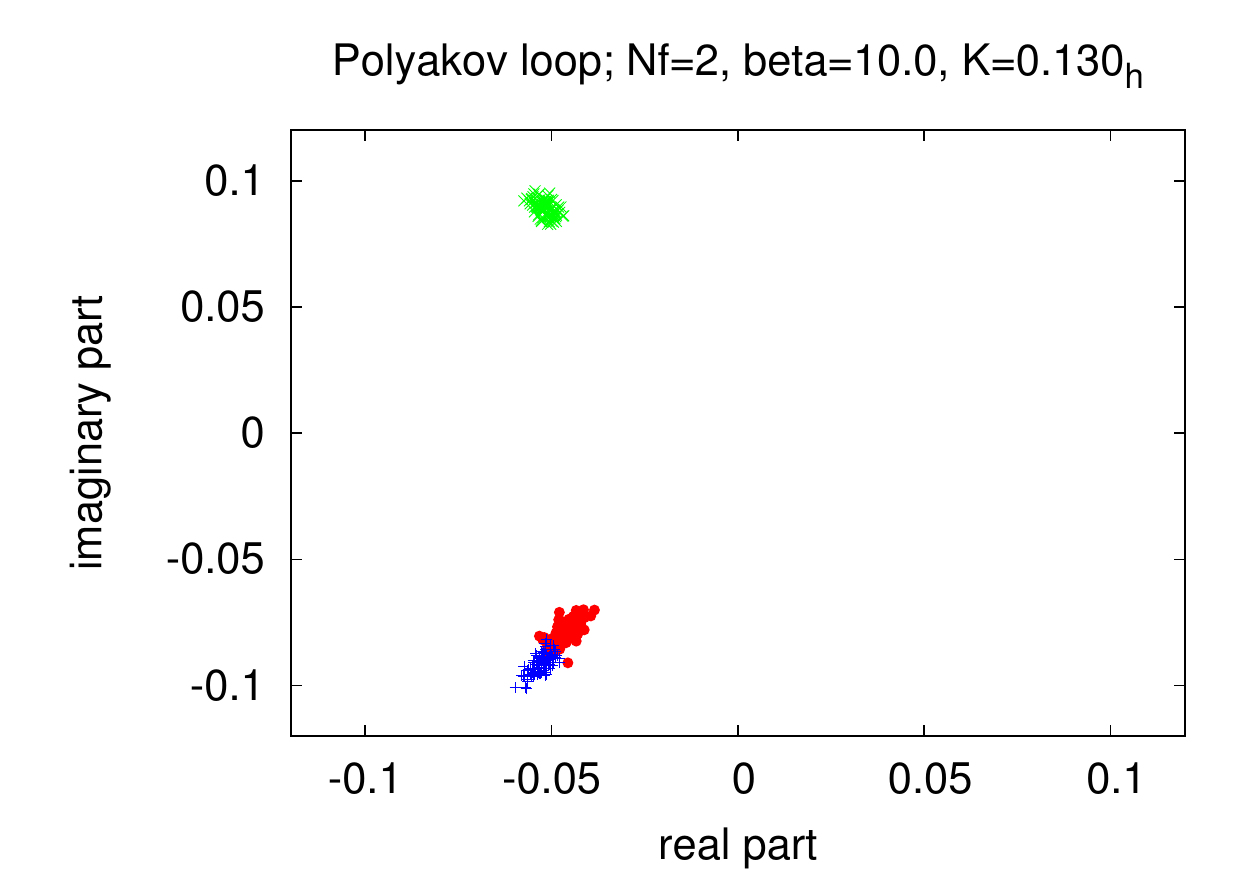}
    \hspace{1cm}
            \includegraphics[width=8.0cm]{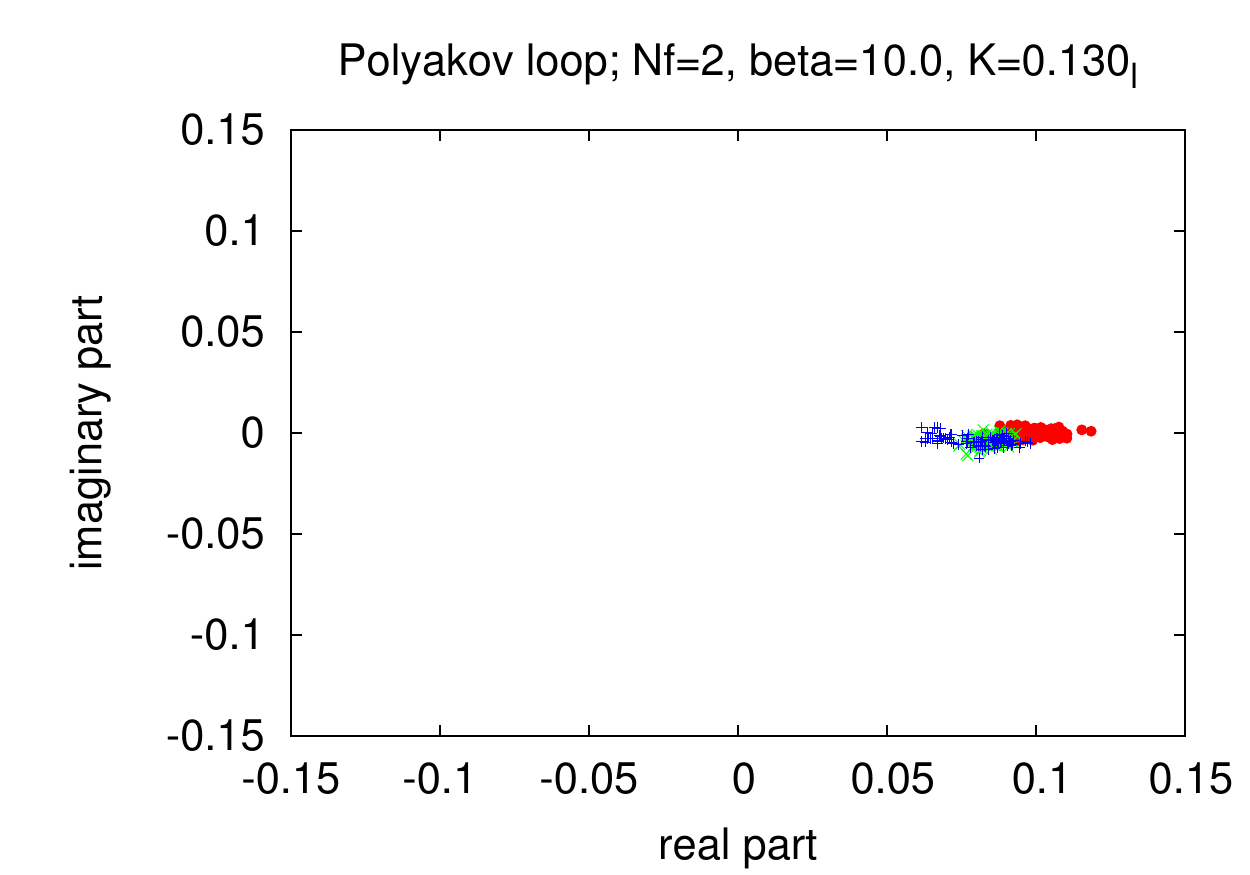}
                                       \caption{(color online) The scattered plots of Polyakov loops in the $x$, $y$ and $z$ directions overlaid; 
        both for $N_f=2$ at $\beta=10.0$ and $K=0.130$: (left) from larger $K$ and (right) from smaller $K$.}
                                             \label{complex_k130}
\end{figure*}

With all the theoretical arguments above, in the remaining part of this paper, we do perform the modest check of our proposal of ``conformal field theories with an IR cutoff" at $T/T_c > 1$ in the case $N_f=2$ on a finite lattice with fixed size  $16^3\times 64$.
Our final goal is the verification of the conjecture of the conformal theory with an IR cutoff 
for the case of the thermodynamical limit in the flat space. 
On the other hand, in order to verify the idea of the conformal theories with an IR cutoff from the temporal propagators, we need a lattice with large $N_t$.Therefore we take the lattice size $16^3\times 64$.
This choice of lattice size may not be appropriate  
for the investigation of the thermodynamical limit in the flat space.

Since our theoretical argument only relies on the vanishing beta function and  the existence of IR cutoff (either by temporal one or spatial one),
this lattice size does not spoil our objective to investigate qualitatively the behavior of propagators. 
If we could confirm our concepts of conformal field theory with IR cutoff on this size lattice,
we would be able to naturally conjecture that our proposal  will be realized on a larger spatial lattice such as $256^3\times 64$.
When we are able to perform simulations on such a large lattice,
we will make the spectral decomposition of $G_H(t)$ by using e.g.
the maximal entropy method (MEM)\cite{hatsuda2004} to compare with experiment.
However this is beyond the scope of this article.

We make the same ansatz  based on the RG argument as in the large $N_f$ within the conformal window.
The behavior of $G_H(t)$ qualitatively differs depending  on whether the quark mass is smaller than the critical mass or not.

When the theory is in the relatively heavy quark region, it decays exponentially at large $t$ as
\begin{equation} G_H(t) = c_H \, \exp(-m_H t)\label{exp}.\end{equation}

On the other hand, 
in the ``conformal region" 
 defined by
\begin{equation} 
m_H   \leq c \,  \Lambda_{\mathrm{IR}} 
\label{critical mass}
\end{equation}
where $c$ is a constant of order 1, the propagator  $G_H(t)$ behaves at large $t$  as
 \begin{equation}
G_H(t) = \tilde{c}_H\\ \frac {\exp(-\tilde{m}_Ht)}{t^{\alpha_H}},
\label{yukawa}
 \end{equation}
which is a power-law corrected Yukawa-type decaying form instead of the exponential decaying form observed when $m_H > c \,  \Lambda_{\mathrm{IR}}$. 
At finite temperature, eqs. (3) and (5) are valid only approximately due to the finiteness of the $t$ range.
A more rigorous way would be to make the spectral decomposition of $G_H(t)$. We postulate, however, the qualitative features of transition should be seen by applying eqs. (3) and (5) on a finite lattice.

Now let us discuss the results of our numerical calculations
in the $N_f=2$ case  for several sets of parameters.

First we identify the chiral transition around $K=0.151$ at $\beta=6.0$ on a lattice $16^3\times 64$
by the ''on-Kc method'' in ref.~\cite{Stand26}. 
Since the transition is second order or weak first order (see the discussion below), it is not easy to ping down the chiral phase transition by monitoring the number of iterations of CG inversion. However, for the purpose of this article a very  precise value
is not necessary. We safely state $5.9 < \beta_c < 6.1$ and $0.150 < K_c < 0.152$.

Then we will verify the conformal region and the conformal behavior at $\beta \ge 6.0$ with small  quark mass.
We choose
the following values of $\beta's$\,:
$\beta=6.5, 7.0, 8.0, 10.0$ and  $15.0$. 
We show the parameters for simulations and the numerical results in Table 1.

\begin{table*}[thb]
\caption{Numerical results for $N_f=2$:
"s" in the second column represents the initial status; the continuation from the lower $K$ (l) or from the higher $K(h)$;
the third column is the number of trajectories for measurement;
the 4th column is the plaquette value;
the $m_q$ in the 5th column is the quark mass;
 $m$ in the 6th and 7th columns are the
mass of $PS$ and $V$ channels in the case of the exponential  decay defined in Eq.~\ref{exp};
the $\tilde{m}$  in 8th and 10th columns and $\alpha$ in the 9th and 11th columns are, respectively,  the "mass"  and the exponent of $PS$ and $V$ channels in the case of the yukawa type  decay defined in Eq.~\ref{yukawa}.}
\begin{tabular}{lrrllllllll}
\hline
\hline
 & \multicolumn{4}{c}{$N_f=2$}   &&&&&& \\
\hline
$K$ & s& $N_{tra}$&  plaq & $m_q$ & $m_\pi$ & $m_V$ & $\tilde{m_{\pi}}$ &$\alpha_\pi$ &$\tilde{m_{V}}$ & $\alpha_V$\\
\hline
&&&&&&$\beta=5.9$&&&&\\
\hline
0.152 & l & 1000 & 0.602192(18) &   0.0332(1) &      0.3280(60) &     0.4492(59) &-&-&-&-\\
\hline
&&&&&&$\beta=6.5$&&&&\\
\hline
0.110 & l &  500 &   0.596366(17)  & 1.6932(23)  &    2.1372(15)  &    2.1398(15)  &-&-&-&-\\
0.145 & l   & 1000 &  0.648107(13)  &   0.0587(2)   &    0.4249(45)  &    0.4414(49) &-&-&-&-\\
0.1455 & l  & 600 &  0.648321(13) &   0.0465(3)  &     0.4112(56)  &    0.4194(70)  &-&-&-&-\\
0.146 & l & 1000 & 0.648546(14) &   0.0337(3)  &     -         &      -       &                0.371(9) &       0.71(8)     &    0.371(12)   &    0.98(14)\\
0.1465 & l & 1000 &  0.648799(14) &   0.0213(4) &      -       &        -      &                 0.286(19) &      0.73(19) &       0.279(14) &      1.08(26)\\
0.147 & l & 1000  &  0.649046(14)  &   0.0083(2) &      -       &        -      &                 0.295(16)  &     1.00(16) &       0.286(6)   &     1.41(20)\\
\hline
&&&&&&$\beta=7.0$&&&&\\
\hline
0.142 &  l &  700 &  0.678445(09)  & 0.0592(3)     &  -   &            -     &                  0.386(13)   &    0.74(15)    &     0.402(11)  &   0.66(10)\\
0.143  & l  &  500 &   0.678788(10) &  0.0333(4) &      - &              -   &                 0.360(13)  &     0.69(22)   &      0.356(10) &     0.94(23)\\
0.144 &  l  &  600 &  0.679108(16)  & 0.0074(2)    &   -    &           -      &                 0.354(14)     &    1.02(14)   &  0.320(14)  &   1.87(18)\\
\hline
&&&&&&$\beta=8.0$&&&&\\
\hline
0.139 & l &  700 & 0.725022(14)  &  0.0345(2) &      - &              -   &                    0.318(12)  &     0.97(14)   &     0.299(12)   &    1.41(21)\\
0.140 & l &  800 &  0.725140(91)  &  0.0084(1) &       - &              -   &                    0.376(7)  &     1.02(7)  &     0.403(6)  &     0.67(9)\\
\hline
&&&&&&$\beta=10.0$&&&&\\
\hline
0.110 &  l &  600 &  0.783954(05) &  0.8644(2)  &     1.3959(5)   &    1.3953(5) &-&-&-&-\\
0.125 & l &  600  &  0.784657(10) &  0.3046(1)  &     0.6518(16) &     0.6520(16) &-&-&-&-\\
0.130 &  l &  700 &   0.785016(08) &   0.1626(1) &       0.3887(5) &       0.3907(7) &-&-&-&-\\
0.130 & h &  900 &   0.785036(11) &  0.1676(1)  &     -        &       -       &                0.495(11)   &    1.40(11)  &       0.498(10)  &      1.32(11)\\
0.135 &  l & 1000 &   0.785549(08) &  0.0280(2) &      -       &        -      &                 0.372(69)  &     1.11(6)  &       0.373(3)  &      1.14(10)\\
\hline
&&&&&&$\beta=15.0$&&&&\\
\hline
0.130 & l & 1000 &  0.860880(03) &  0.0455(1)  &     -   &            -    &                   0.385(55)  &     1.21(4)  &       0.3972(6)  &     1.00(9)\\
\hline
\end{tabular}
\end{table*}

We take several values of $K$ for each $\beta$ in such a way that the quark masses $m_q's$ take values $0.00 \le m_q \le 0.10.$ 
We also verify when mass is heavy, the propagator behaves as an exponentially decaying form.
It may be worthwhile to stress that when $\beta < 6.0$,  the propagator behaves as an exponentially decaying form even for very small quark mass.

The algorithm we employ is the blocked HMC algorithm \cite{Hayakawa:2010gm}.
We choose run-parameters in such a way that the acceptance of the global metropolis test is about $70\%.$
The statistics are 1,000 MD trajectories for thermalization and 1,000 MD trajectories or 500 MD trajectories for the measurement.
We estimate the errors by the jack-knife method
with a bin size corresponding to 100 HMC trajectories.

We define the effective mass $m_H(t)$  by 
$$\frac{\cosh(m_H(t)(t-N_t/2))}{\cosh(m_H(t)(t+1-N_t/2))}=\frac{G_H(t)}{G_H(t+1)}\ .$$

FIG.~\ref{exp-yukawa} shows 
the $t$ dependence of the effective mass 
for the PS channel  at $\beta=6.5$:
on the left panel for a relatively heavy quark mass $K=0.140 \, (m_q=0.18(1))$ and on the right panel for a light quark mass $K=0.146 \, (m_q=0.034(1))$.
Three types of symbols represent three types of source-sink;  the local-sink local-source (squares)
local-sink (quark-anti-quark) doubly-exponentially-smeared-source of a radius 5 lattice units  (circles) and 
local-sink doubly-wall-source (triangles).
On the left panel, we see the clear plateau of the effective mass at $t=20\sim 31$.
On the other hand, on the right panel, we see the effective mass is slowly decreasing without no plateau up to $t=31$,
suggesting the power-law correction.
The power-law corrected fit for the local-local data with the fitting range $t=[15:31]$  with $\alpha_H=0.71(8)$ reproduces the data very well.

When $\beta=5.9$ the effective mass plot also shows the plateau even for a very small quark mass 
$m_q=0.0046$ ($K=0.153$).

These results are consistent with the conjecture: 
When the quark mass is light $m_q=0.034(1)$, it is in the conformal region at $\beta=6.5$, while when it is relatively heavy $m_q=0.18(1)$ it is outside of the conformal region. 
On the other hand, for $\beta < \beta^c=6.0$, there is no conformal region.

The RG argument predicts the boundary of the conformal region and outside of the conformal region is first order,
and in the large $N_f$ within the conformal window this prediction was confirmed in the $N_f=7$ case~\cite{coll}.

We show a typical example of first order transition in FIg.~\ref{effm_k130}; 
the effective mass plot of the two states at the same parameter at $\beta=10.0$ with $K=0.130, (m_q=0.16(1))$.
The state on the left was obtained from a configuration at larger $K$ as an initial state, while the state on the right was from smaller $K$.
These two states persist for more than one thousand trajectories.
We see clearly the difference between the two: we have a plateau on the right and a power-law corrected one on the left.
There is also a large difference in the effective mass: 0.39(1) vs. 0.57(1).
This result shows that the transition is first order, as predicted by RG argument.

We also show the scattered plots of the Polyakov loops in the complex plane in Figs.\ref{complex_k130}. 
On the left panel the arguments are $\pm 2/3\pi$ and the magnitudes are $\simeq 0.05$ which is non-trivial $Z(3)$ twisted one, while on the right panel the arguments are $0$ and the magnitudes are $\sim 0.06 \sim 0.12$: It is characteristic in the deconfining region.

We have calculated the effective vacuum energy on a finite lattice at $\beta=\infty$ in the one-loop approximation\cite{coll-full}. It turns out that the lowest energy state is the non-trivial $Z(3)$ twisted vacuum which takes the arguments $\pm 2/3\pi$, when the quark mass is light. The numerical results above are consistent with the analytic result. Thus the transition across the boundary is a first order transition between different vacua.

\begin{figure}[htb]
\includegraphics [width=7.5cm]{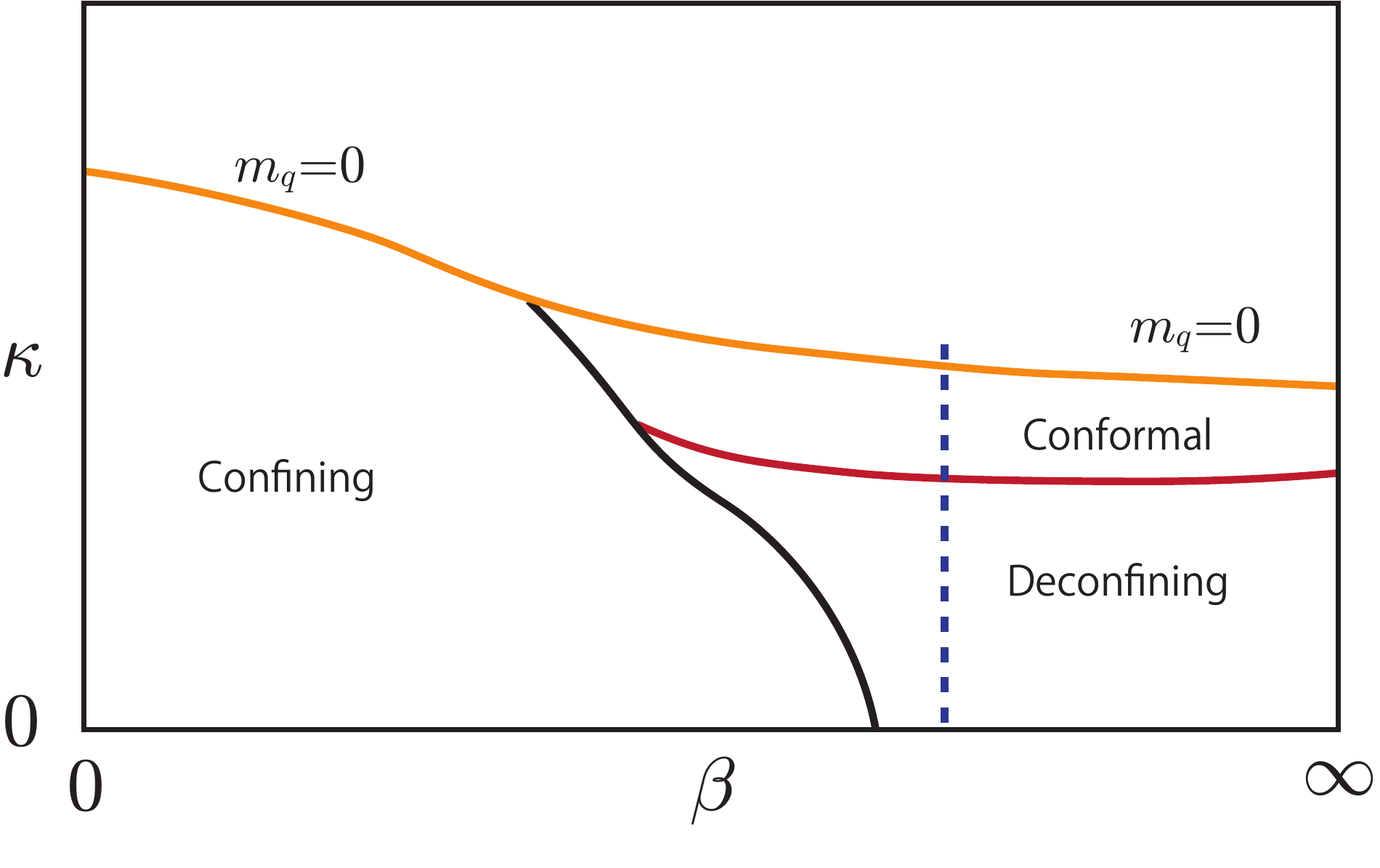}
 \caption{(color online) The phase diagram on a finite lattice:  the solid line toward the quench QCD $K=0$ ($\beta=6.7$) is
the boundary between the deconfining and confining region on the $16^3\times 64$ lattice. The dashed line represents the simulation line $\beta=10.0$. }
\label{conformal_region_3}
\end{figure}

Based on our numerical simulations,
we propose the phase structure as shown in Fig.~\ref{conformal_region_3}. Our result supports this phase structure.
For all cases with $0.0 \le m_q \le 0.10$ at various $\beta$'s, we observe the similar power corrected behavior as
shown on the right panel in Fig. 2.
Thus we have verified our conjecture on the existence of the conformal region and the conformal behavior in the conformal region.

Now we would like to discuss the implication of the existence of the IR fixed point in high temperature QCD.
In this article we concentrate on the issue of
the order of the chiral phase transition in the $N_f=2$ case. 
Our key observation is the existence of an IR fixed point at $T>T_c$.
 We stress that the reasoning for the existence can be justified even in the  thermodynamic limit.

Pisarski and Wilczek \cite{pisarski}
mapped $N_f=2$ QCD at high temperature 
to the three dimensional sigma model and 
pointed out that if $U_A(1)$ symmetry is not recovered at the chiral transition temperature,
the chiral phase transition of QCD in the $N_f=2$ case is 2nd order 
with exponents of the  three dimensional $O(4)$ sigma model.

The $O(4)$ scaling relation was first tested for staggered quarks~\cite{karsch94}.
For the Wilson quarks it was shown
that
the chiral condensate satisfies remarkably  the $O(4)$  scaling relation,
with the RG improved gauge action and the Wilson quark action \cite{iwa1997}  and 
with the same gauge action and the clover-improved Wilson quark action \cite{cppacs2001} (See, for example, Fig.6 in ref.~\cite{iwa1997}).
It was also shown for staggered quarks the scaling relation is satisfied 
 in the $N_f=2 + 1$ case \cite{staggered}, extending the region from $T/T_c >1$ adopted in
 \cite{iwa1997} and  \cite{cppacs2001}
 to the region including $T/T_c<1$.
 These results imply the transition is second order.
 
 However, recently, it was shown that the expectation value of 
 the chiral susceptibility $\chi_{\pi} -\chi_{\sigma}$ is
 zero~\cite{aoki2012} in thermodynamic limit
 when the $SU(2)$ chiral symmetry is recovered under the assumptions we will discuss below. This is consistent with that the  $U_{A}(1)$ symmetry is recovered, which
implies the transition is 1st order according to \cite{pisarski}. 
Apparently the  two conclusions are in contradiction.

Here we revisit this issue with the new insight of conformal field theories with an IR cutoff.
It is assumed in ref.~\cite{aoki2012}  that
the vacuum expectation value of mass-independent observable is an analytic function of 
$m_q^2$, if the chiral symmetry is restored.
However, in the conformal region
the propagator of a meson behaves as eq.(3) and the relation between the $m_H$ and the $m_q$ is given by the hyper-scaling relation~\cite{miransky} 
$$m_H = c \, m_q^{1/(1+\gamma*)},$$
with $\gamma^*$ the anomalous mass dimension.
This anomalous scaling implies $m_H$  is not analytic in terms of $m_q^2$ and  the analyticity assumption does not hold.
It should be noted that the Ward -Takahashi Identities
in~\cite{aoki2012} are proved in the thermodynamic limit and therefore
the numerical verification of the hyper-scaling in the limit will be decisive.
We stress however that the hyper-scaling is theoretically derived with
the condition of the existence of the IR fixed point and
multiplicative renormalization of $m_q$. We believe that this
violation of the analyticity assumption resolves the apparent
discrepancy as also mentioned in~\cite{aoki2012} as a viable
possibility.


In the article in preparation
we hope to discuss other physical implications of conformal field theories with an IR cutoff realized at high temperature QCD which are not discussed in this article,
together with the discussion on the relation between  large $N_f$ QCD within the conformal window and small $N_f$ QCD at high temperature.

\section*{Acknowledgments}
We would like to thank 
S. Aoki, H. Fukaya, K. Kanaya, T Hatsuda and Y. Taniguchi,
for useful discussion.
The calculations were performed with HA-PACS computer at CCS, University of Tsukuba and SR16000
at KEK. We would like to thank members of CCS and KEK for their strong support for this work.


\end{document}